\documentclass[a4paper]{jpconf}
\usepackage{graphicx}
\begin{document}
\title{Cu hyperfine coupling constants of HgBa$_{2}$Ca$_{}$Cu$_{2}$O$_{6+\delta}$}

\author{Yutaka Itoh$^1$, Takato Machi$^2$ and Ayako Yamamoto$^3$}
\address{$^1$Department of Physics, Graduate School of Science, Kyoto Sangyo University, Kamigamo-Motoyama, Kika-ku, Kyoto 603-8555, Japan}   
\address{$^2$AIST Tsukuba East, Research Institute for Energy Conservation, 1-2-1 Namiki, Tsukuba, Ibaraki 305-8564, Japan} 
\address{$^3$Graduate School of Engineering and Science, Shibaura Institute of Technology, 3-7-5 Toyosu, Koto-ku, Tokyo 135-8548, Japan}

\ead{yitoh@cc.kyoto-su.ac.jp}

\begin{abstract}
We estimated the ratios of $^{63}$Cu hyperfine coupling constants in the double-layer high-$T_\mathrm{c}$ superconductor HgBa$_2$CaCu$_2$O$_{6+\delta}$ from the anisotropies in Cu nuclear spin-lattice relaxation rates and spin Knight shifts to study the nature of the ultraslow fluctuations causing the $T_2$ anomaly in the Cu nuclear spin-echo decay. The ultraslow fluctuations may come from uniform magnetic fluctuations spread around the wave vector $q$ = 0, otherwise the electric origins.
\end{abstract}

\section{Introduction}
Spin polarized neutron scattering experiments indicate the emergence of an intra-unit-cell (IUC) $q$ = 0 magnetic moments in the pseudogap states of the high-$T_\mathrm{c}$ cuprate superconductors, while no NMR and $\mu$SR experiment indicates any static ordering of local magnetic moments~\cite{Sidis}. 
The IUC moments are associated with the loop current ordered state~\cite{Varma}.
Recently discovered ultraslow fluctuations in the pseudogap states of HgBa$_2$CaCu$_2$O$_{6+\delta}$ (Hg1212) via $^{63}$Cu nuclear spin-echo decay experiments~\cite{ItohULS} might reconcile an issue on the IUC moments. 
No wipeout effect on NMR spectra is characteristic of the ultraslow fluctuations of Hg1212, in contrast to the spin-charge stripe orderings~\cite{stripe1,stripe2}. 

Knowledge of the hyperfine coupling constants helps us to clarify the nature of the local field fluctuations in NMR measurements~\cite{MR}.    
In this paper, we report the estimation of the $^{63}$Cu hyperfine coupling constants in the double-CuO$_2$-layer high-$T_\mathrm{c}$ superconductors Hg1212 from the anisotropies in $^{63}$Cu nuclear spin-lattice relaxation rates and spin Knight shifts~\cite{ItohHg1212}, and discuss the nature of the ultraslow fluctuations~\cite{ItohULS}. 

\section{Estimations of $^{63}$Cu hyperfine coupling constants} 
The $^{63}$Cu hyperfine coupling parameters ($A_{cc}^{hf}$ and $A_{ab}^{hf}$) consist of the anisotropic on-site $A_{cc}$ (the $c$ axis component) and $A_{ab}$ (the $ab$ plane component) due to the 3$d$ electrons and the isotropic supertransferred component $B (> 0)$~\cite{MR}.  
The ratios of the individual components in the three coupling constants can be estimated from the anisotropy data~\cite{ItohHg1212} of the $^{63}$Cu Knight shifts ($^{63}K_{cc}$ and $^{63}K_{ab}$) and the $^{63}$Cu nuclear spin-lattice relaxation rates [(1/$T_1$)$_{cc}$ and (1/$T_1$)$_{ab}$] via the antiferromagnetic dynamical spin susceptibility~\cite{MR,MMP,MPS,Imai,Itoh2}.
The subscript indices of $cc$ or $ab$ of $^{63}K_{}$ and 1/$T_1$ indicate the direction of a static magnetic field applied along the $c$ axis or in the $ab$ plane. 
The procedure to estimate the coupling constant ratios is shown below. 

\subsection{$^{63}$Cu Knight shifts}
The $^{63}$Cu Knight shifts $^{63}K_{cc,ab}$ at a magnetic field along the $c$ axis or in the $ab$ plane are the sum of the spin shift $K_{spin}$ and the orbital shift $K_{orb}$ as $^{63}K_{cc,ab} = K_{spin}^{cc,ab}(T) + K_{orb}^{cc,ab}$.
The spin shift is $K_{spin}^{cc,ab}(T) = A_{cc,ab}^{hf}\chi_{s}^{cc,ab}(T)$ with the hyperfine coupling parameters $A_{cc,ab}^{hf}$ and the uniform spin susceptibility $\chi_{s}^{cc,ab}(T)$. 
For a temperature-dependent isotropic spin susceptibility $\chi_{s}^{cc} = \chi_{s}^{ab}$, the ratio $\Delta K_{spin}^{cc}/\Delta K_{spin}^{ab} = (dK_{spin}^{cc}/dT)/(dK_{spin}^{ab}/dT)$ is equal to the ratio $A_{cc}^{hf}/A_{ab}^{hf}$. 

\begin{figure}[t]
\begin{minipage}{16pc}
\includegraphics[width=16.5pc]{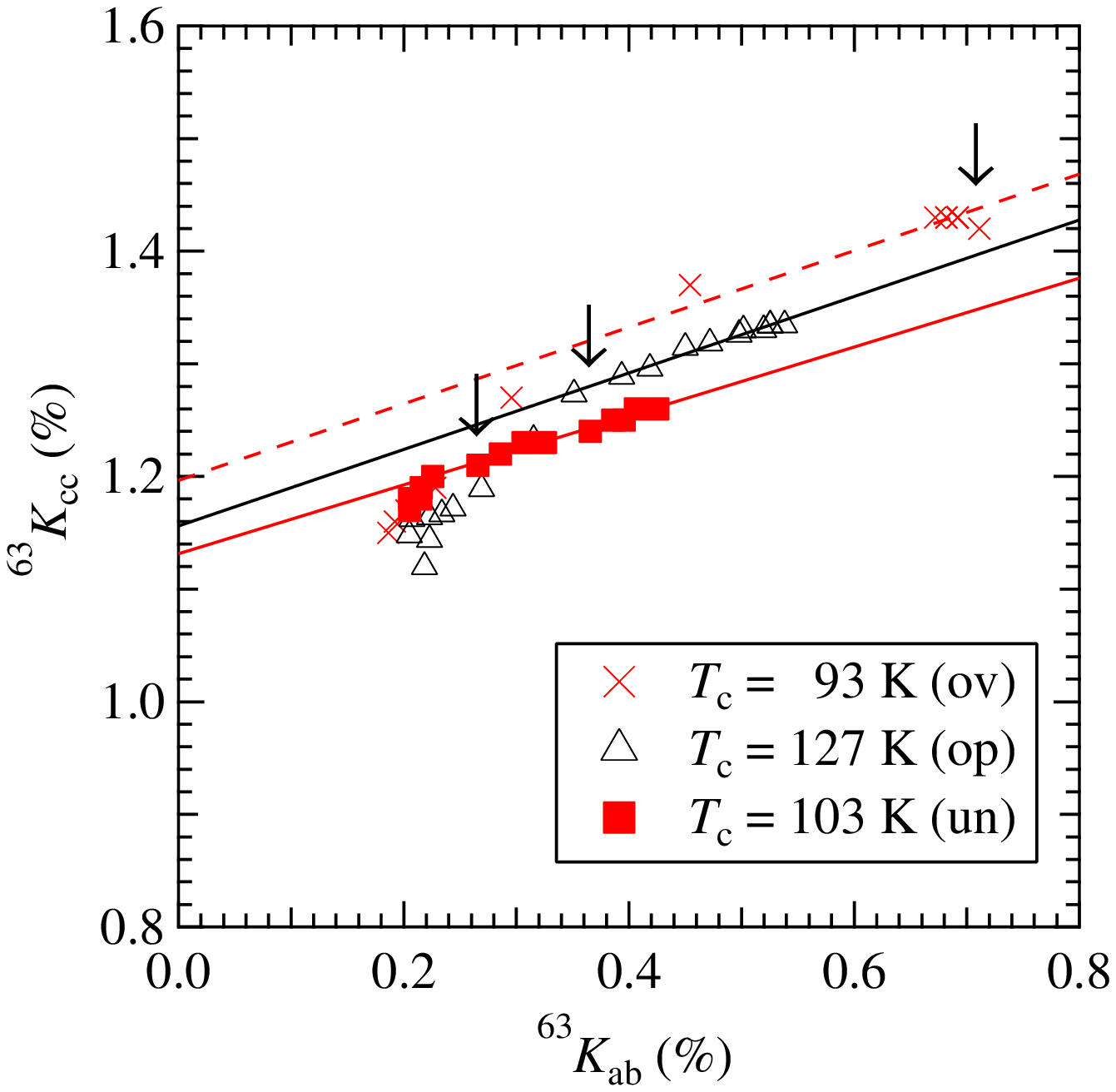}
\caption{\label{F1}
$^{63}K_{cc}$ versus $^{63}K_{ab}$ with temperature as an implicit parameter for Hg1212 from underdoped to overdoped~\cite{ItohHg1212}. The arrows indicate $T_\mathrm{c}$.   
}
\end{minipage}\hspace{2pc}%
\begin{minipage}{16pc}
\includegraphics[width=16.0pc]{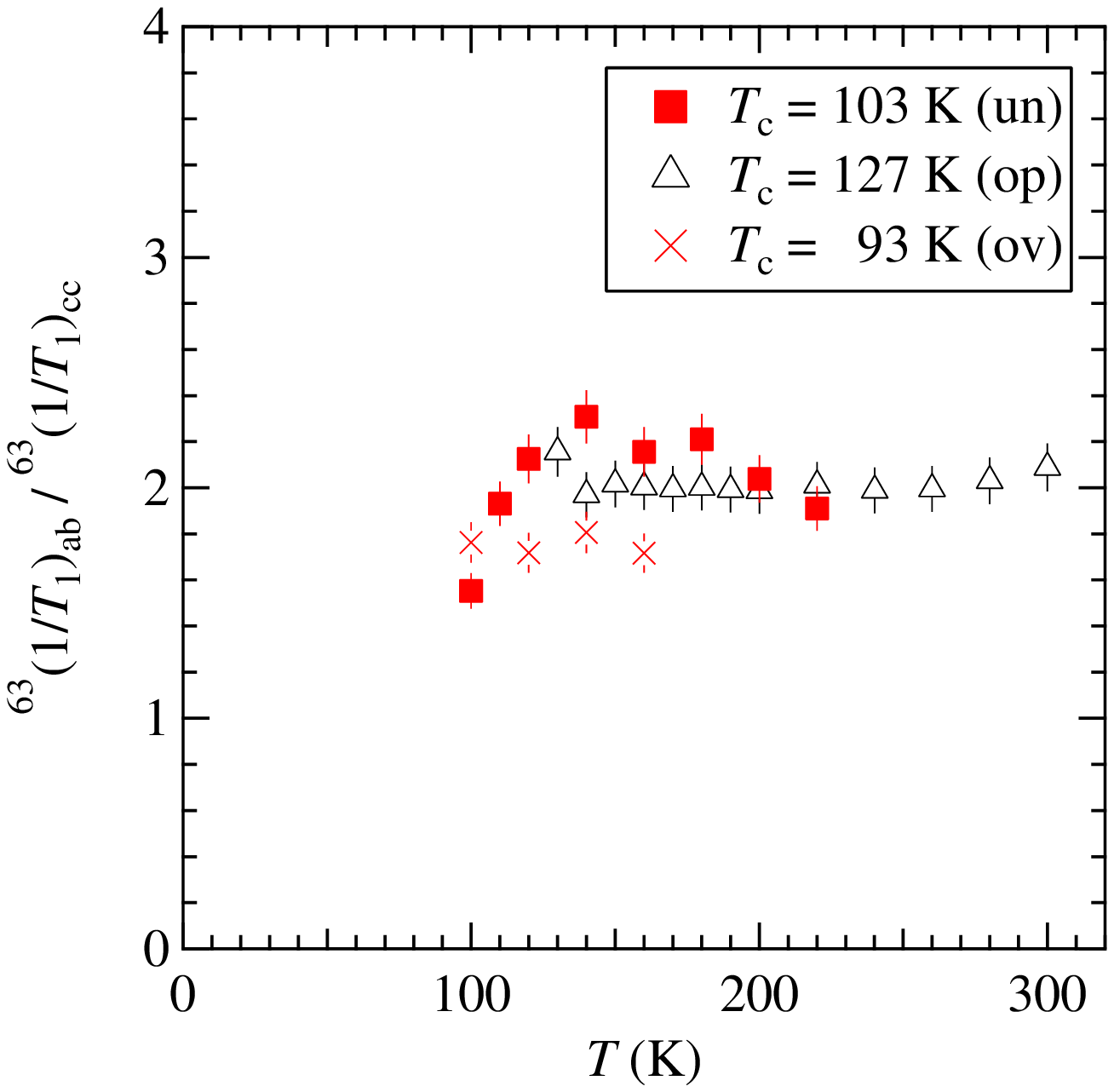}
\caption{\label{F2}
Anisotropy ratio $^{63}$(1/$T_{1}$)$_{ab}$/$^{63}$(1/$T_{1}$)$_{cc}$ against temperature for Hg1212 from underdoped to overdoped~\cite{ItohHg1212}.
}
\end{minipage} 
\end{figure}

Figure~\ref{F1} shows $^{63}K_{cc}$ plotted against $^{63}K_{ab}$ with temperature as an implicit parameter for Hg1212 from underdoped to overdoped, which are adopted from~\cite{ItohHg1212}. 
The sold straight lines are the least-squares fitting results. 
The dashed straight line for overdoped Hg1212 is a visual guide with assuming the same slope as the optimally doped Hg1212. 
The straight lines show nearly parallel shift. 
Since the orbital shifts of $K_{orb}^{cc}$ = 1.14$-$1.16 $\%$ and $K_{orb}^{ab}$ = 0.19$-$0.20 $\%$ are estimated below $T_\mathrm{c}$, then the parallel shift indicates a constant spin component above $T_\mathrm{c}$. 
Similar parallel shift is found in the single crystal NMR for HgBa$_2$CuO$_{4+\delta}$~\cite{Jurugen}. 

An easy plane magnetic anisotropy causes such a constant spin component in the paramagnetic spin susceptibility~\cite{Shimizu}. 
The anisotropic superexchange interaction in the $S$ = 1/2 XXZ Heisenberg Hamiltonian yields the easy plane anisotropy in the paramagnetic state~\cite{Hanzawa,XXZ}. 
The optimally hole doping makes the anisotropy weak~\cite{Shimizu,Terasaki}.
Although the multicomponents in the spin susceptibility are suggested from the anisotropic spin Knight shifts~\cite{Jurugen,JHaase}, 
we believe that the doped superconductors with a single spin component can show a finite anisotropy and that the constant spin component does not impede a single spin component analysis to estimate the Cu hyperfine coupling constants.  

The $^{63}$Cu hyperfine coupling parameters $A_{cc}^{hf}$ and $A_{ab}^{hf}$ are expressed by $A_{cc}$, $A_{ab}$, and $B$ as
 $A_{cc}^{hf} = A_{cc}+4B$ and $A_{ab}^{hf} = A_{ab}+4B$~\cite{MR,MMP,Takigawa}.
Then, the anisotropy ratio $r_u$ of the temperature-dependent $K_{s}^{cc}$ and $K_{s}^{ab}$ is given by  
\begin{eqnarray}
r_{u} \equiv {\Delta K_{s}^{cc} \over {\Delta K_{s}^{ab}}} = {A_{cc}+4B \over {A_{ab}+4B}}.  
\label{ru}
\end{eqnarray}
Figure~\ref{F1} shows $r_u$ = 0.31 for the underdoped and 0.34 for the optimally doped samples. 
The value of $r_u$ = 0.34 is assumed for the overdoped sample. 

\subsection{$^{63}$Cu nuclear spin-lattice relaxation rate}
Figure~\ref{F2} shows the ratio of $^{63}$(1/$T_{1}$)$_{ab}$/$^{63}$(1/$T_{1}$)$_{cc}$ plotted against temperature for Hg1212 from underdoped to overdoped (adopted from Ref.~\cite{ItohHg1212}).
The anisotropy ratio $r_\mathrm{AF}$ of (1/$T_1$)$_{ab}$ and (1/$T_1$)$_{cc}$ is given by 
\begin{eqnarray}
r_\mathrm{AF} \equiv {(1/T_{1})_{ab} \over {(1/T_{1})_{cc}}} \approx {1\over 2}\biggl\{1+\biggl({A_{cc}-4B \over {A_{ab}-4B}}\biggr)^2 \biggr\}, 
\label{rAF}
\end{eqnarray}
for the leading term of the enhanced antiferromagnetic susceptibility~\cite{Itoh2}. 
For convenience, we introduce an alternative parameter of $r_{A} = \sqrt{2r_\mathrm{AF}-1}$.  
We adopted the values of $r_{AF}$ ($r_A$) = 2.3 (1.90), 2.0 (1.73), and 1.8 (1.61) from underdoped to overdoped (figure~\ref{F2}) to estimate the coupling constant ratios. 

\subsection{$^{63}$Cu hyperfine coupling constant ratios}
From the constraints of $A_{cc}$ $<$ 0~\cite{Takigawa} and $A_{ab}/4B$ $<$ 1 on~(\ref{ru}) and~(\ref{rAF}),  
we obtain the expressions of the ratios of the $^{63}$Cu hyperfine coupling constants, 
\begin{eqnarray} 
{A_{cc} \over {4B}} \approx  - {{r_{A}+r_{u} - 2r_{A}r_{u}} \over {r_{A}-r_{u}}},\\  
{A_{ab} \over {4B}} \approx {{r_{A}+r_{u} - 2} \over {r_{A}-r_{u}}}, 
\label{AccAab}
\end{eqnarray}
and then
\begin{eqnarray} 
{A_{ab} \over {A_{cc}}} \approx  - {{r_{A}+r_{u} - 2} \over {r_{A}+r_{u} - 2r_{A}r_{u}}}. 
\label{Accab}
\end{eqnarray}
Thus, (3)-(5) with a set of $r_u$ and $r_A$ enable us to estimate the ratios of the $^{63}$Cu hyperfine coupling constants. 

\begin{table}[b]
\caption{\label{table1}
Anisotropies ($r_u$ and $r_{AF}$) of $^{63}K_s$ and $^{63}T_1$, and the ratios of $^{63}$Cu hyperfine coupling constants ($A_{cc}$, $A_{ab}$, $B$)
for underdoped (un), optimally doped (op) and overdoped (ov) Hg1212. 
$T_\mathrm{c}$ is in kelvin.
The value of $r_u$ for overdoped Hg1212 is assumed after the optimally doped value in figure 1.
} 
\begin{center}
\lineup
\begin{tabular}{*{7}{l}}
\br                              
$\0\0$&$\0T_\mathrm{c}$&$\0\0r_u$&$r_{AF}$&\m$A_{cc}/4B$&\m$A_{ab}/4B$&\m$A_{ab}/A_{cc}$\cr 
\mr
\0\0un&103&$\0$0.31&2.3&$\0-0.65$&$\0+0.13$&$\0-0.20$\cr
\0\0op&127&$\0$0.34&2.0&$\0-0.64$&$\0+0.05$&$\0-0.078$\cr 
\0\0ov&$\0$93&$\0$0.34&1.8&$\0-0.67$&$\0-0.04$&$\0+0.058$\cr  
\br
\end{tabular}
\end{center}
\end{table}

Table~\ref{table1} shows the estimated ratios of $A_{cc}/4B$, $A_{ab}/4B$ and $A_{ab}/A_{cc}$ for Hg1212 from (3)-(5) with the experimental $r_u$ and $r_{AF}$ in figures~\ref{F1} and~\ref{F2}. 
The on-site coupling ratio $A_{ab}/A_{cc}$ depends on the hole concentration in Hg1212. 

The 3$d(x^2 - y^2)$ orbital electron of Cu$^{2+}$ in the tetragonal crystal field produces the on-site hyperfine fields. 
The ratio $A_{ab}/A_{cc}$ is expressed as
\begin{eqnarray} 
{A_{ab} \over {A_{cc}}} \approx  - {{-\kappa + {{2}\over{7}} - {{11}\over{7}}\gamma} \over {-\kappa - {{4}\over{7}} - {{62}\over{7}}\gamma}}, 
\label{Accab}
\end{eqnarray}
where $\kappa$($>$ 0) is  the core polarization parameter, 2/7 and $-$4/7 are the spin-dipole field coefficients, and $\gamma$($<$ 0) is the spin-orbit coupling parameter~\cite{MPS,Pennington,Bleaney}.
The empirical values of $\kappa$ = 0.25 and 0.325 were estimated for Cu$^{2+}$ ions in the dilute copper salts~\cite{Bleaney}. The first-principles cluster calculations give $\kappa$ = 0.289 for the density functional $<1/r^3>$ and 0.455 for the Hartree-Fock $<1/r^3>$ in La$_2$CuO$_4$~\cite{LCO}.  
The value of $\kappa$ = 0.41 is found in CuGeO$_3$~\cite{MItoh}.
For Hg1212, $A_{ab}/A_{cc}$ in Table 1 through (6) leads to the core polarization parameter $\kappa$ = 0.265 (un), 0.315 (op) and 0.387 (ov), assuming $\gamma$ = $-$0.044~\cite{MPS,Pennington}. 

\subsection{$^{63}$Cu hyperfine coupling constants of HgBa$_2$CuO$_{4+\delta}$ and Hg1212}
Let us show the $^{63}$Cu hyperfine coupling constants of the optimally doped single-CuO$_2$-layer superconductor HgBa$_2$CuO$_{4+\delta}$ ($T_\mathrm{c}$ = 98 K). 
From the uniform spin susceptibility $\chi_s$ = 1.47$\times$10$^{-4}$ emu/mole-f.u.~\cite{Nova} and the in-plane $^{63}$Cu $K_{spin}^{ab}$ = 0.48 $\%$~\cite{ItohHg1201}, we estimated the in-plane $^{63}$Cu hyperfine coupling parameter $A_{ab}^{hf}$ = $A_{ab} + 4B$ = ($N_\mathrm{A}\mu_\mathrm{B}/\chi_s)K_{spin}^{ab}$ = 182 kOe/$\mu_\mathrm{B}$ for HgBa$_2$CuO$_{4+\delta}$ 
($N_\mathrm{A}$ is Avogardro's number and $\mu_{B}$ is the Bohr magneton).
Substituting $r_u$ = 0.53 and $r_{AF}$ = 1.8~\cite{ItohHg1201} into (3)-(5) and using $A_{ab}$ + 4$B$ = 182 kOe/$\mu_\mathrm{B}$, we obtained the values of
\begin{center}
$A_{cc}$ = $-$ 65, $A_{ab}$ = 21, and $B$ = 40 kOe/$\mu_B$
\end{center}
for the optimally doped HgBa$_2$CuO$_{4+\delta}$. 
\begin{table}[b]
\caption{\label{tabone}
$^{63}$Cu hyperfine coupling constants in units of kOe/$\mu_\mathrm{B}$ for underdoped (un), optimally doped (op) and overdoped (ov) Hg1212, assuming $A_{ab} + 4B$ = 182 kOe/$\mu_\mathrm{B}$.
} 
\begin{center}
\lineup
\begin{tabular}{*{4}{l}}
\br                              
$\0\0$&\m$A_{cc}$&$A_{ab}$&$B$\cr 
\mr
\0\0un&$-105$&$+21$&$40$\cr
\0\0op&$-111$&$+8.7$&$43$\cr 
\0\0ov&$-127$&$-7.4$&$47$\cr  
\br
\end{tabular}
\end{center}
\end{table}
\begin{figure}[h]
\includegraphics[width=17pc]{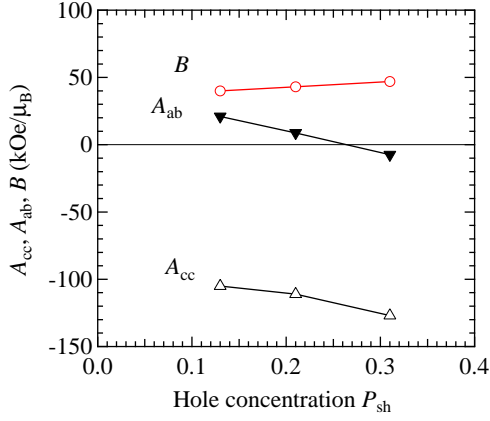}\hspace{2pc}%
\begin{minipage}[b]{17pc}\caption{\label{F3}
$^{63}$Cu hyperfine coupling constants $A_{cc}$, $A_{ab}$, and $B$ plotted against hole concentration $P_{sh}$ for Hg1212.
The solid curves are visual guides. 
}
\end{minipage}
\end{figure} 

By adopting $A_{ab} + 4B$ = 182 kOe/$\mu_\mathrm{B}$ for Hg1212 after HgBa$_2$CuO$_{4+\delta}$, we estimated the individual components of $A_{cc}$, $A_{ab}$, and $B$ (Table~\ref{tabone}). 
Figure~\ref{F3} shows $A_{cc}$, $A_{ab}$, and $B$ (Table~\ref{tabone}) plotted against the hole concentration $P_{sh}$~\cite{ItohHg1212} for Hg1212.
In Table~\ref{tabone} and figure~\ref{F3}, with increase in the hole concentration, the absolute value of the negative $A_{cc}$ increases, $A_{ab}$ shows a sign change, and the $B$ term slightly increases.

The reported $B$ term is in the range from 36 to 155 kOe/$\mu_\mathrm{B}$ in the other cuprate superconductors
~\cite{Imai,Kitaoka,Kambe,Ishida,Mukuda}, 
assuming {\it a priori} the fixed values of $A_{cc}$ = $-$170 and $A_{ab}$ = 37 kOe/$\mu_\mathrm{B}$~\cite{Kitaoka,Ishida,Mukuda}. 
The cation-cation supertransferred hyperfine field $B$ between 3$d$ and 4$s$ orbitals depends on the strength of the $p$-$d$ covalent bond parameter~\cite{Huang}. 
The doping dependent $B$ term in Table 2 indicates the development of the covalency with the hole concentration in Hg1212. 
 
\section{Local field fluctuations in $^{63}$Cu nuclear spin-echo decay rate 1/$T_{2L}$}
\subsection{$^{63}$Cu nuclear spin-echo decay rate 1/$T_{2L}$}
Figures 4(a)-4(c) show the $^{63}$Cu nuclear spin-echo decay rates (1/$T_\mathrm{2L}$)$_{ab,cc}$'s for Hg1212 from underdoped (a), optimally doped (b) and overdoped (c)~\cite{ItohULS}. 
The notations conform to those in~\cite{ItohULS}.
The enhancements in (1/$T_\mathrm{2L}$)$_{ab,~cc}$ at 220$-$240 K indicate the ultraslow fluctuations~\cite{ItohULS}.
The peak temperature of (1/$T_\mathrm{2L}$)$_{cc}$ is nearly independent of the doping level, 
but the enhancement is suppressed by overdoping. 

Figure 4(d) shows the anisotropy ratio of the local field fluctuations $\Delta J_{cc}/\Delta J_{ab}$ $\equiv$ [(1/$T_\mathrm{2L}$)$_{cc}$ $-$ (1/$T_\mathrm{2R}$)$_{cc}$]/[(1/$T_\mathrm{2L}$)$_{ab}$ $-$ (1/$T_\mathrm{2R}$)$_{ab}$] derived from 1/$T_\mathrm{2L}$ and 1/$T_1$ (Redfield's 1/$T_\mathrm{2R}$)~\cite{ItohULS}. 
$\Delta J_{\gamma\gamma}$ ($\gamma\gamma$ = $cc$ and $ab$) expresses the additional fluctuations causing the enhancement in 1/$T_\mathrm{2L}$.
One should note that $\Delta J_{cc}/\Delta J_{ab} <$ 1 is characteristic of the ultraslow fluctuations.  

\begin{figure}[b]
\includegraphics[width=23pc]{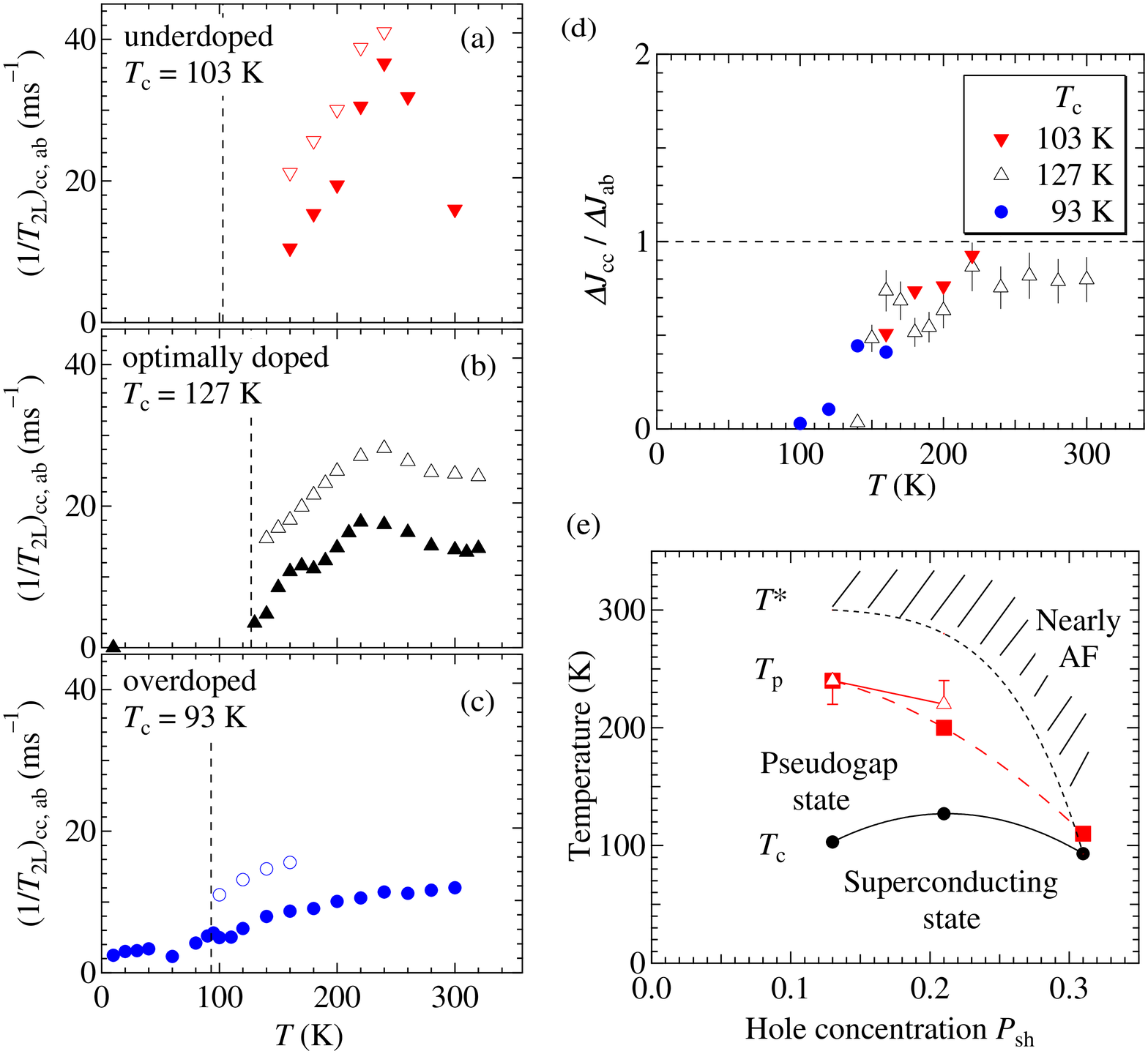}\hspace{0pc}%
\begin{minipage}[b]{14.5pc}\caption{\label{F4}
(a)$-$(c) $T$ dependences of (1/$T_\mathrm{2L}$)$_{cc}$ (closed symbols) and  (1/$T_\mathrm{2L}$)$_{ab}$ (open symbols) from underdoped to overdoped Hg1212~\cite{ItohULS}. Each dashed line indicates $T_\mathrm{c}$.  
(d) $T$ dependences of $\Delta J_{cc}/\Delta J_{ab}$~\cite{ItohULS}.
(e) Phase diagram of Hg1212: $T_\mathrm{c}$ (closed circles), the pseudo spin-gap temperature defined by the maximum temperature of 1/$T_1T$~\cite{ItohHg1212} (closed squares), $T_\mathrm{p}$ defined by the peak temperature of (1/$T_\mathrm{2L}$)$_{cc}$~\cite{ItohULS} (open triangles) against hole concentration $P_{sh}$. 
The dotted curve with a shaded region is a visual guide for the onset temperature $T^*$ of decrease in $^{63}$Cu Knight shift~\cite{ItohHg1212}. 
Nearly AF stands for the Curie-Weiss law in Cu 1/$T_1T$~\cite{ItohHg1212}.
}
\end{minipage}
\end{figure} 
Figure 4(e) shows the phase diagram of Hg1212, where 
the superconducting transition temperature $T_\mathrm{c}$, the pseudo spin-gap temperature defined by the maximum temperature of 1/$T_1T$, $T_\mathrm{p}$ defined by the peak temperature of (1/$T_\mathrm{2L}$)$_{cc}$, and the onset temperature $T^*$ of the decrease in the Cu Knight shift are plotted against the hole concentration $P_{sh}$ in Cu$^{2+P_{sh}}$~\cite{ItohULS,ItohHg1212}.    
With hole doping, $T^*$ decreases, while $T_\mathrm{p}$ is nearly independent of the hole concentration $P_{sh}$.
The ultraslow fluctuations emerge in the underdoped regime and diminish in the overdoped regime. 
\begin{figure}[b]
\begin{minipage}{16pc}
\includegraphics[width=16pc]{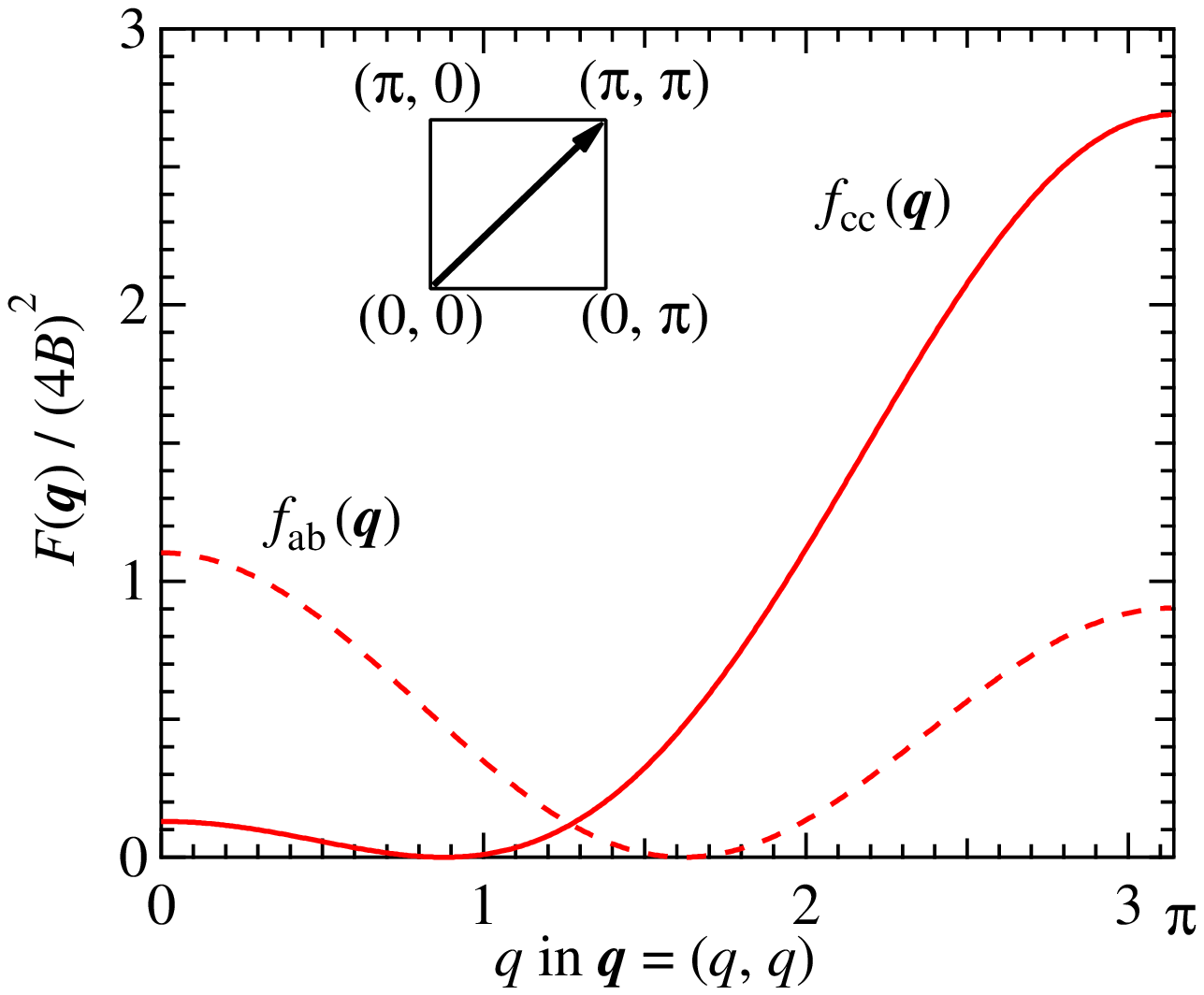}
\caption{\label{F5}
$^{63}$Cu hyperfine coupling form factors $f_{cc}$({\mbox{\boldmath $q$}}) and $f_{ab}$({\mbox{\boldmath $q$}})
as functions of $q$ in the wave vector {\mbox{\boldmath $q$}} = ($q$, $q$) [(0, 0) $\rightarrow$ ($\pi$, $\pi$)]. 
Inset shows the diagonal in the first Brillouin zone. 
}
\end{minipage}\hspace{2pc}%
\begin{minipage}{16pc}
\includegraphics[width=16pc]{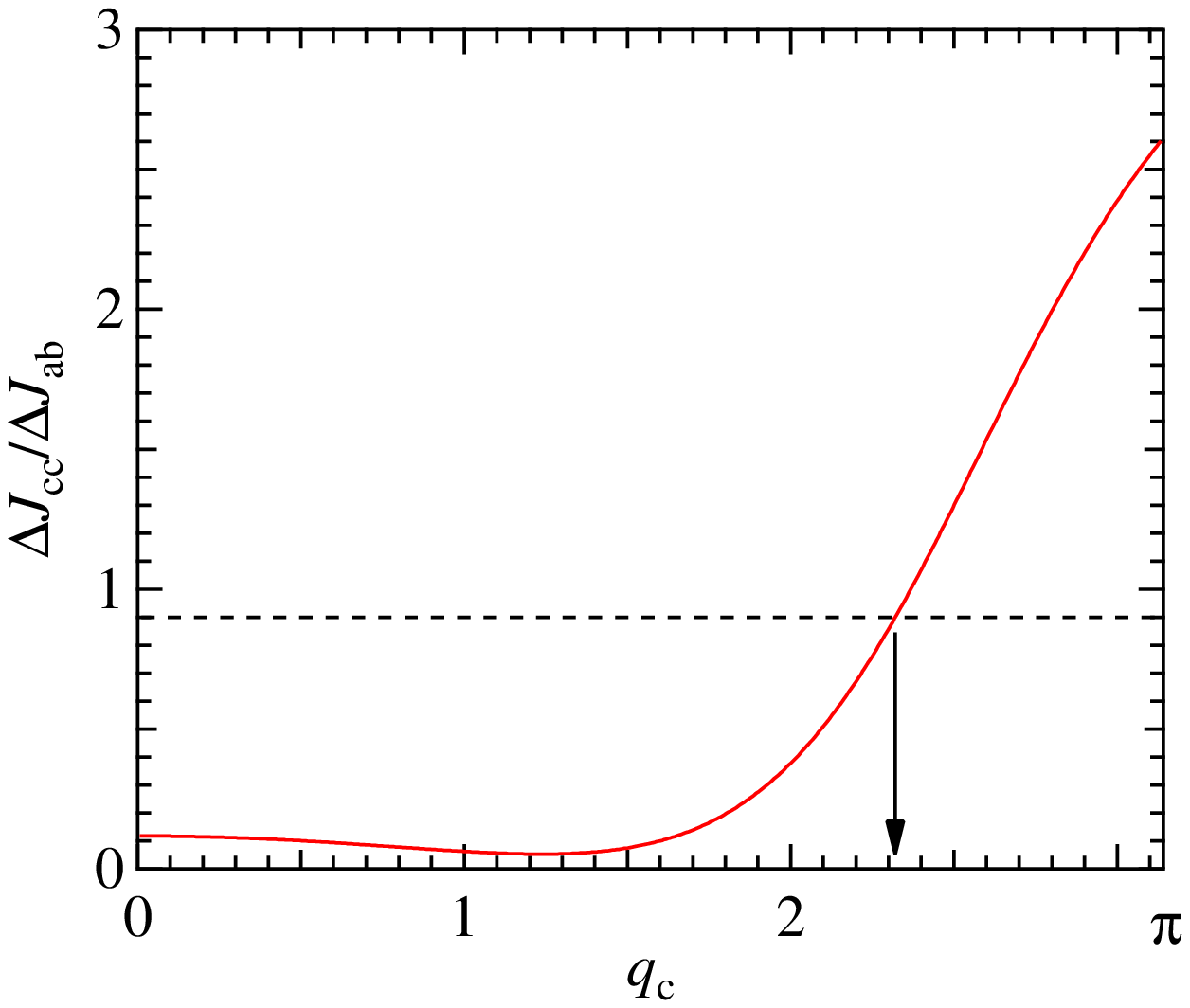}
\caption{\label{F6}
$\Delta J_{cc}/\Delta J_{ab}$ as a function of the cut off $q_c$ in a toy model upon a static spin susceptibility $\chi^{\prime}({\mbox{\boldmath $q$}})$ = $\chi_0^{\prime}$ ($\left|q_{x,y}\right| < q_c$) and 0 ($\left|q_{x,y}\right| > q_c$). 
Experimental constraint leads to $q_c$ $<$ 2.3. 
}
\end{minipage} 
\end{figure}

\subsection{Local field fluctuations}
Local field fluctuations of $J_{ab}$ ($B\perp$c axis) and $J_{cc}$ ($B\parallel$c axis) causing the nuclear spin relaxations of $T_1$ and $T_2$ are defined by
\begin{eqnarray} 
J_{\gamma\gamma} = \sum_{\mbox{\boldmath $q$}}F_{\gamma\gamma}({\mbox{\boldmath $q$}})S_{}({\mbox{\boldmath $q$}}, \nu_n),\\
F_{\gamma\gamma}({\mbox{\boldmath $q$}}) \equiv (4B)^2f_{\gamma\gamma}({\mbox{\boldmath $q$}}) = [A_{\gamma\gamma} + 2B\{\mathrm{cos}(q_x)+\mathrm{cos}(q_y)\}]^2,
\label{Fab}
\end{eqnarray}
where $\gamma\gamma$ = $ab$ and $cc$, and $\nu_n$ is an NMR frequency~\cite{ItohULS}. 
The electron spin-spin correlation function $S_{}$({\mbox{\boldmath $q$}}, $\nu$) (a frequency $\nu$) is related to the dynamical spin susceptibility $\chi^{\prime\prime}$({\mbox{\boldmath $q$}}, $\nu$) through the fluctuation-dissipation theorem. 
$F_{\gamma\gamma}$({\mbox{\boldmath $q$}}) is called the form factor of the wave vector {\mbox{\boldmath $q$} dependent hyperfine coupling constant, whose filtering effects in the {\mbox{\boldmath $q$} space play a significant role in the anisotropy and the site differentiation on NMR~\cite{MMP,MPS,Imai}. 
$\Delta J_{\gamma\gamma}$ expresses the additional fluctuations to $J_{\gamma\gamma}$~\cite{ItohULS}.

Figure~\ref{F5} shows the $q$ dependence of $f_{ab,~cc}$({\mbox{\boldmath $q$}}) for Hg1212 along the diagonal {\mbox{\boldmath $q$}} = ($q$, $q$) in the first Brillouin zone, using the estimated coupling constant ratios in Table~\ref{table1}. 
Since $f_{cc}$($\pi$, $\pi$) $>$ $f_{ab}$($\pi$, $\pi$), the antiferromagnetic spin fluctuation $\chi^{\prime}$({\mbox{\boldmath $q$}}) localized around {\mbox{\boldmath $q$}} = [$\pi$, $\pi$] leads to the anisotropy $J_{cc}/J_{ab}\sim$ 3 in contrast to the experimental ratio $\Delta J_{cc}/\Delta J_{ab}<$ 1 in figure 4(d)~\cite{ItohULS}.  
$\Delta J_{\gamma\gamma}$ expresses the development of the ultraslow fluctuations~\cite{ItohULS}.
Thus, the antiferromagnetic fluctuations are excluded from the ultraslow fluctuations. 

Let us assume a toy model of $\chi^{\prime}$({\mbox{\boldmath $q$}}, $\nu_n$) = $\chi^{\prime}_0\Theta(q_c - \left|q_x\right|)\Theta(q_c - \left|q_y\right|)$  ($\Theta(x)$ is the Heaviside step function). 
$q_c$ is a cut-off wave number.
$\chi^{\prime}({\mbox{\boldmath $q$}})$ $\propto$ $S$({\mbox{\boldmath $q$}}, $\nu_n$) takes a constant value $\chi^{\prime}_0$ over $\left|q_{x,y}\right| < q_c$. 
For this toy model, the ratio $\Delta J_{cc}/\Delta J_{ab}$ is calculated as
\begin{eqnarray} 
{\Delta J_{cc} \over {\Delta J_{ab}}} = {{\sum_{\left| q_{x, y}\right | < q_c}f_{cc}({\mbox{\boldmath $q$}})} \over {\sum_{\left| q_{x, y}\right | < q_c}f_{ab}({\mbox{\boldmath $q$}})}}. 
\label{Jccab}
\end{eqnarray}
Figure~\ref{F6} shows the numerical $\Delta J_{cc}/\Delta J_{ab}$ as a function of $q_c$. 
The experimental $\Delta J_{cc}/\Delta J_{ab}$ $<$ 0.9 in figure 4(d) imposes on the function in figure~\ref{F6} and then leads to $q_c$ $<$ 2.3. 
The magnetic ultraslow fluctuations must be confined within $q_c$ $<$ 2.3. 
If the magnetic ultraslow fluctuations have the easy plane anisotropy, the upper limit of the cut-off value $q_c$ will be smaller than 2.3. 
Thus, we obtained a model constraint on the magnetic ultraslow fluctuations, using the anisotropic hyperfine coupling constants. 

Although the step function $\chi^{\prime}({\mbox{\boldmath $q$}})$ with $q_c$ $<$ 2.3 is not localized at $q$ = 0, it is parallel to the IUC $q$ = 0 magnetic moments observed by the spin polarized neutron scattering method~\cite{Sidis}. 
The ultraslow fluctuations may be associated with the IUC $q$ = 0 magnetic moments. 
However, if the enhancement in 1/$T_\mathrm{2L}$ is due to quadrupole fluctuations, one should explore the alternative fluctuations of charge or lattice for the electric ultraslow fluctuations.

\section{Conculsions}
The systematic hole doping dependences of the $^{63}$Cu hyperfine coupling constants ($A_{cc}$, $A_{ab}$ and $B$) 
were found for Hg1212 from underdoped to overdoped.
A model constraint on the magnetic ultraslow fluctuations in Hg1212 was derived from the anisotropy ratios of the $^{63}$Cu hyperfine coupling constants. 
The model expresses the magnetic fluctuations spread around $q$ = 0. 
Possible electric ultraslow fluctuations causing the $T_2$ anomaly remain to be explored.  

\ack 
We thank Jun Kikuchi for fruitful discussions on the hyperfine coupling constants.

\section*{References}

\smallskip
 


\medskip
\begin{thebibliography}{9}
\bibitem{Sidis}Bourges P and Sidis Y 2011 {\it C. R. Physique} {\bf 12} 461
\bibitem{Varma}Varma C M 2014 {\it J. Phys.: Condens. Matter} {\bf 26} 505701
\bibitem{ItohULS}Itoh Y, Machi T and  Yamamoto A 2017 {\it Phys. Rev.} B {\bf 95} 094501
\bibitem{stripe1}Singer P M, Hunt  A W, Cederstr\"{o}m  A F and Imai T 1999 {\it Phys. Rev.} B {\bf 60} 15345 
\bibitem{stripe2}Hunt  A W, Singer P M, Cederstr\"{o}m  A F and Imai T 2001 {\it Phys. Rev.} B {\bf 64} 134525 
\bibitem{MR}Mila F and Rice T M 1989 {\it Physica} C {\bf 157} 561 
\bibitem{ItohHg1212}Itoh Y, Tokiwa-Yamamoto A, Machi T and Tanabe K 1998 {\it J. Phys. Soc. Jpn.} {\bf 67} 2212
\bibitem{MMP}Millis A J, Monien H and Pines D 1990 {\it Phys. Rev.} B {\bf42} 167
\bibitem{MPS}Monien H, Pines D and Slichter C P 1990 {\it Phys. Rev.} B {\bf 41} 11120  
\bibitem{Imai}Imai T 1990 {\it J. Phys. Soc. Jpn.} {\bf 59} 2508 
\bibitem{Itoh2}Itoh Y, Hayashi A and Ueda Y 1995 {\it J. Phys. Soc. Jpn.} {\bf 64} 3074  
\bibitem{Jurugen}Rybicki D, Kohlrautz J, Haase J, Greven M, Zhao X, Chan M K, Dorow C J and Veit M J 2015 {\it Phys. Rev.} B {\bf 92} 081115 
\bibitem{Shimizu}Shimizu T, Aoki H, Yasuoka H, Tsuda T, Ueda Y, Yoshimura K and Kosuge K 1993 {\it J. Phys. Soc. Jpn.} {\bf 62} 3710
\bibitem{Hanzawa}Hanzawa K 1994 {\it J. Phys. Soc. Jpn.} {\bf 63} 264  
\bibitem{XXZ}Okabe Y and Kikuchi M 1988 {\it J. Phys. Soc. Jpn.} {\bf 57} 4751
\bibitem{Terasaki}Terasaki I, Hase M, Maeda A, Uchinokura K, Kimura T, Kishio K, Tanaka I and Kojima H 1992 {\it Physica} C {\bf 193} 365 
\bibitem{JHaase}Haase J, Jurkutat M, Kohlrautz J 2017 {\it Condense. Matter} {\bf 2} 16
\bibitem{Takigawa}Takigawa M, Hammel P C, Heffner R H, Fisk Z, Smith J L and Schwarz R B 1989 {\it Phys. Rev.} B {\bf 39}  300
\bibitem{Pennington}Pennington C H, Durand  D J, Slichter C P, Rice J P, Bukowski  E D and Ginsberg D M 1989 {\it Phys. Rev.} B {\bf 39} 2902
\bibitem{Bleaney}Bleaney B, Bowers  K D and Pryce M H L 1955 {\it Proc. Roy. Soc. London}, Ser. A {\bf 228} 166
\bibitem{LCO}H\"{u}sser P, Suter H U, Stoll E P and Meier P F 2000 {\it Phys. Rev.} B {\bf 61} 1567 
\bibitem{MItoh}Itoh M, Sugahara M, Yamauchi T and Ueda Y 1996 {\it Phys. Rev.} B {\bf 53} 11606
\bibitem{Nova}Itoh Y, Machi T and Yamamoto A [arXiv:2091155] 
\bibitem{ItohHg1201}Itoh Y, Machi T, Adachi S, Fukuoka A, Tanabe K and Yasuoka H 1998 {\it J. Phys. Soc. Jpn.} {\bf 67} 312
\bibitem{Kitaoka}Kitaoka Y, Fujiwara K, Ishida K, Asayama K, Shimakawa Y, Manako T and Kubo Y 1991 {\it Physica} C {\bf 179} 107
\bibitem{Kambe}Kambe S, Yasuoka H, Hayashi A and Ueda Y 1993 {\it Phys. Rev.} B {\bf 47} 2825
\bibitem{Ishida}Ishida K, Kitaoka Y, Asayama K, Kadowaki K and Mochiku T 1994 {\it J. Phys. Soc. Jpn.} {\bf 63} 1104
\bibitem{Mukuda}Shimizu S, Iwai S, Tabata S-I, Mukuda H, Kitaoka Y, Shirage P M, Kito H and Iyo A 2011 {\it Phys. Rev.} B {\bf 83} 144523
\bibitem{Huang}Huang N L, Orbach R, \v{S}im\'{a}nek E, Owen J and Taylor D R 1967 {\it Phys. Rev.} {\bf 156} 383
\end{thebibliography}
\end{document}